\documentclass[a4paper,fleqn,usenatbib]{mnras}
\usepackage{newtxtext,newtxmath}
\usepackage[T1]{fontenc}
\usepackage{ae,aecompl}
\usepackage{graphicx}	
\usepackage{amsmath}	
\usepackage{amssymb}	

\newcommand{\kms}{km\,s$^{-1}$}

\title[SN~2010jl: dusty line-emitting shell]
{Type IIn SN~2010jl: probing dusty line-emitting shell}

\author[N. N. Chugai]{
Nikolai N. Chugai,$^{1}$\thanks{E-mail: nchugai@inasan.ru}
\\
Institute of Astronomy Russian Academy of Science, Moscow, Russia\\
}

\date{Accepted XXX. Received YYY; in original form ZZZ}

\pubyear{2016}

\begin{document}
\label{firstpage}
\pagerange{\pageref{firstpage}--\pageref{lastpage}}
\maketitle

\begin{abstract}
I explore signatures of a possible dust formation in the late SN~2010jl that could be  
imprinted in the line blueshift and the radius evolution of the dusty 
infrared-emitting shell.
I propose a simple model that permits one to reproduce emission lines of 
blueshifted hydrogen and helium emission lines. The model suggests that 
the hydrogen emission originates primarily from shocked fragmented circumstellar clumps partially 
obscured by the absorbing cool dense shell and by unshocked ejecta. In the He\,1.083 $\mu$m line 
on day 178 this component is significantly weaker compared to broad component from unshocked ejecta 
that is obscured by the absorprion produced by ejecta itself.
Simulations of late time ($t > 400$ d) H$\alpha$ suggest that, apart from the dust in the cool 
dense shell, a significant amount 
of dust must form in the unshocked supernova ejecta. The supernova radius 
predicted by the interaction model coincides with the radius of the dusty shell recovered from 
late time (> 460 days) infrared data, which strongly support that infrared radiation indeed originates from 
supernova. The ejecta dust is presumably locked in opaque blobs. 
\end{abstract}

\begin{keywords}
supernovae -- individual -- SN 2010jl
\end{keywords}



\section{Introduction} 
\label{sec:intro}

The luminous supernova SN~2010jl interacting with a dense circumstellar matter (CSM) 
is a special case among SNe~IIn due to a large volume of 
observational data accumulated in different bands including  X-ray \citep{Chandra2012,Ofek2014,Chandra2015}, 
optical \citep{Stoll2011,Zhang2012,Smith2012,Ofek2014,Fransson2014,Gall2014,Borish2015,Jencson2016}, 
infrared (IR) \citep{Andrews2011,Fransson2014,Gall2014,Borish2015,Dwek2017}, and radio \citep{Chandra2015}.
This provide us with an opportunity to get deeper insight into complicated phenomena 
accompanied the ejecta interaction  with a dense CSM.
Of particular interest is the suspected dust formation in the postshock cool dense 
shell (CDS) indicated by the late time hot IR excess \citep{Maeda2013,Gall2014} combined with the 
line blueshift \citep{Smith2012}. The estimated amount of the dust is 
of $\approx 10^{-3}\,M_{\odot}$ after about 500 d \citep{Maeda2013,Gall2014,Sarangi2018}.

The possibility of the dust formation in the CDS of SNe~IIn was discussed earlier for SN~1998S 
 \citep{Pozzo2004}, SN~2006jc  \citep{Smith2008,Mattila2008,Chugai2009}, SN~2005ip \citep{Smith2009},
 SN~2007od \citep{Andrews2010}.
Yet there are open questions that may cast doubts on this conjecture. Among them is 
the issue of so called blueshifted Lorentz-like profiles (narrow core plus broad wings).
The point is that, if the dusty line-emitting shell is composed of the 
 CDS material only, one would expect then to find a blueshifted  broad boxy line profile 
 because in this case the velocity dispersion of the line-emitting material is small 
 compared to the expansion velocity. This issue was emphasised by \citet{Fransson2014} to argue in favour of the CS dust. 
 The concept of the external dust however leaves blueshifted Lorentz-like profiles 
 unexplained. Since the conjecture on the dusty CDS and the line blueshift seems to be closely linked 
 the origin of the Lorentz-like profile requires an explanation before a confident conclusion on 
 the dust location could be made. Furthermore, accepting the idea of the dust formation in the 
 CDS we have not so far answers to a straightforward questions concerning the site and velocity distribution 
 of the line-emitting gas and the dust distribution.
 It is noteworthy that, apart from the CDS, the dust in SNe~IIn might form also in unshocked ejecta 
in the same way as it does in SNe~IIP, e.g., SN~1987A \citep{Lucy1989} and SN~1999em \citep{Elmhamdi2003}. 
Note, the dust formation in the unshocked ejecta of 
SN~IIn has been already proposed in a context of an infrared excess in the SN~2005ip \citep{Fox2010}.
This possibility should be considered as highly probable for SN~2010jl as well.

Below I address two principal questions: (i) what is the origin of blueshifted lines in spectra 
of SN~2010jl; (ii) whether the late time SN radius is consistent with the 
radius of the IR-emitting dust shell. The answer to these questions will permit us 
to present at least a qualitative model that accounts for blueshifted line profiles and late IR emission 
in unique picture lacking at the moment.
I start with an explantion of the origin of Lorentz profile and the blueshift modelling 
(Section \ref{sec:specmod}). This is followed by the modelling of the 
SN/CSM interaction that provides us with an important tool for the probing the relation between 
the SN radius and the radius of the IR-emitting shell (Section \ref{sec:dmod}). 
In the Discussion section the clumpy dust distribution is discussed and the estimate of the 
dust amount is given.

Following \citet{Fransson2014} we adopt the explosion date 2010 October 9 (JD 2455479).

\section{Spectral modelling}
\label{sec:specmod}

\subsection{Preliminaries: Why Lorentz profile?}
\label{sec:preview}

The interaction of spherically-symmetric freely expanding ($v = r/t$) SN ejecta with 
a spherically-symmetric CSM occurs via the formation of two shock waves: forward (radius $R_f$) 
and reverse shock (radius $R_r$) with a shocked SN and CS matter swept-up in a thin dense shell at the 
contact surface (radius $R_0$). (Henceforth radii are measured in units of $R_0 = 1$, if otherwise not 
stated explicitly). In an adiabatic case 
for the typical density distribution of ejecta ($\rho \sim v^{-8}$) and CSM ($\rho \sim r^{-2}$) the 
shock radii ratios are $R_r/R_0 = 0.98$ and $R_f/R_0 = 1.27$ \citep{Chevalier1982b}. In strongly radiative 
regime the postshock layers get narrower, so one can adopt in this case $R_r = R_0$. 
The X-rays from both shocks excite cold unshocked SN 
ejecta and CSM thus giving rise to optical broad lines and narrow lines respectively.
In a realistic situation the picture grow significantly more complicated: the flow between 
forward and reverse shock turns out three-dimensional due to Rayleigh-Taylor (RT) instability of the thin shell 
\citep{Chevalier1982a,CheBlo1995,BloEllis2001} and because of a lumpiness of the CSM. Therefore, in 
reality additional line-emitting sites related to the CDS and shocked CS 
clouds can appear between forward and reverse shock. Due to the non-trivial velocity spectrum and 
highly complicated radiation transfer effects the optical radiation spectrum  of SNe~IIn 
cannot be reliably predicted so far.

\begin{figure}
	\includegraphics[trim=0 0 0 0,width=1\columnwidth]{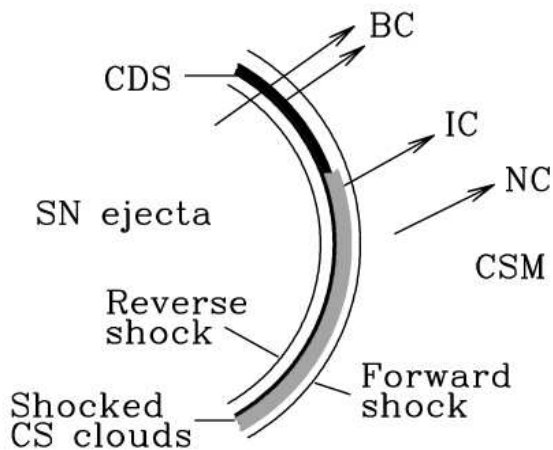}
    \caption{
    Schematic diagram of the SN/CSM interaction showing source location for the narrow, broad 
    and intermediate components involved in the overall line emission. 
    The layer of cold CDS material is partially mixed with the layer of shocked CS clouds ({\em grey}); 
    only a fragment of the latter is shown for clarity.
    }
    \label{fig:cart}
\end{figure}

Emission lines of SNe~IIn at the early stage ($t < 1$ yr) usually (but not always) reveal three 
components:  the narrow component (NC),  broad (BC), and intermediate (IC), --- all originally have been
recognized in the SN~1988Z spectrum by \citet{Filippenko1991}. These components  
can be identified (Figure \ref{fig:cart}) with the CSM, unshocked SN ejecta in combination with the CDS, 
and shocked CS clouds, respectively, \citep{Chugai1994}. 
In SN~2010jl the distinction between BC and IC is not so apparent as in 
SN~1988Z: hydrogen lines, e.g. H$\alpha$,  of SN~2010jl are rather smooth and indeed reminds a "Lorentz" 
profile \citep{Fransson2014}. We will use this term, keeping in mind that 
it has nothing to do with the Lorentz distribution of frequencies of the damping oscillator radiation.
Fortunately, in He\,I 1.083\,$\mu$m line the component ratio of BC/IC is large, so 
both components are easily distinguished \citep{Borish2015}, which keeps safe canonical three-component 
structure of emission lines for SN~2010jl. 

It  should be emphasised that late time Lorentz profiles in SN~2010jl have essentially different 
origin than emission lines with Lorentz profile in early spectra of some SNe~IIn, identified 
with emission of the ionized
dense preshock CSM, where these lines get broadened by the Thomson scattering; 
the phenomenon recognized for SN~1998S \citep{Chugai2001}. 
This type of the profile cannot present at late epochs ($ t \gtrsim 100$ d) since the required  
Thomson optical depth of a preshock CSM must be large ($\tau_{\mbox{\scriptsize T}} \gtrsim 2$) 
that is unattainable at that late time.

The Lorentz profile in the late time SNe~2010jl can be interpreted following a suggestion 
invoked for SN~2006jc (type IIn), where the emission lines have been attributed to shocked fragmented 
CS clouds \citep{Chugai2009}. A typical value of the cloud shock in SNe~IIn 
is $v_c \sim 10^3$\,\kms\ \citep[e.g., SN~1988Z,][]{Chugai1994}, which for the postshock intercloud 
velocity $v_{ic} \approx 6000$\,\kms\ implies the CS cloud density contrast 
with respect to the intercloud density $\chi \approx (v_{ic}/v_c)^2 \sim$ 30-40.
The shocked fragmented CS clouds resides between the forward shock and the CDS 
(Figure \ref{fig:cart}) while their velocities range between the initial cloud shock velocity 
($v_c$) and the final velocity of fragments $v_{max} \approx (0.75...~0.9)v_{ic}$  
for $10 <\chi < 100$ \citep{Klein1994}. Note, the velocity the postshock intecloud gas coincides 
with the CDS velocity $v_{cds}$.
The  cloud fragmentation and fragments acceleration by the incident shock  
is well demonstrated by laser experiments and 3D-hydrodynamic simulations \cite{Klein2003} with  
the relevant physics well understood earlier \citep{Klein1994}. 
 
The case of SN~2006jc indicates that the Lorentz profiles can be reproduced 
with the velocity distribution of the line emissivity  
$j(v) \propto (v_{max} - v)$  \citep{Chugai2009}.
The following toy model illustrates, how such a velocity spectrum could arise.
Consider a steady-state flow of CS clouds into the forward shock with the rate $G$ (clouds/s). 
The cloud in the forward shock experiences crushing by the radiative shock followed by the fragmentation, 
acceleration of fragments, and fragments mixing with ambient intercloud hot gas \citep{Klein1994}.
Let cloud life span in the forward shock be $T$ 
and the cloud survival probability at the age $\tau$ be the linear function of the age
$p(\tau) \propto (1 - \tau/T)$.  Assuming that the fragmented cloud accelerates  
linearly as $v = v_c +a\tau$, which is close to 2D-simulation results \citep{Klein1994},
and taking into account that the clouds age distribution is $dN/d\tau = Gp(\tau)$ one finds 
the velocity distribution of shocked clouds  $dN/dv = (dN/d\tau)(d\tau/dv) \propto (v_{max}-v)$, 
where $v_{max} = v_c + aT$. With the constant specific emissivity (per gramm) we thus come to the 
required velocity distribution of the line emissivity. 
This is only crude illustration and the emissivity velocity distribution may difer from this simple law.

\begin{figure}
	\includegraphics[trim=40 160 0 -20,width=0.95\columnwidth]{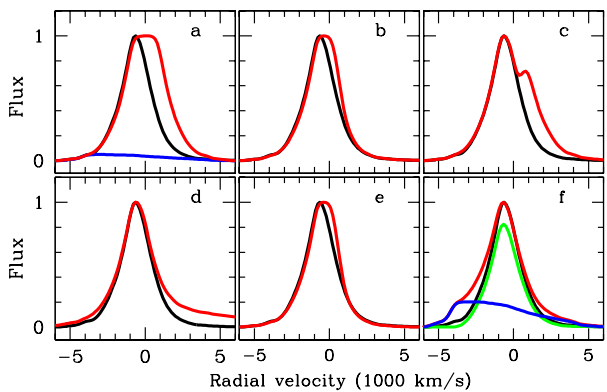}
    \caption{
     Effects of the parameters variation on the H$\alpha$ model profile. 
     Panels from {\bf a} to {\bf f} show the fiducial model mod0 ({\em black}) 
     and models by order from mod1 to mod6
     with changed parameter value ({\em red}) (Table \ref{fig:demo}). 
     In the panel {\bf a} the combined contribution of broad components of the CDS and 
     ejecta in the mod0 is shown for clarity ({\em blue}).
     The panel {\bf f} demonstrates effect of 
     a large contribution of the CDS and unshocked ejecta emission ({\em blue}) 
     while the intermediate component ({\em green}) remains unchanged.
      }
    \label{fig:demo}
\end{figure}

\begin{figure*}
	\includegraphics[trim=60 140 0 -10,  width=0.9\textwidth]{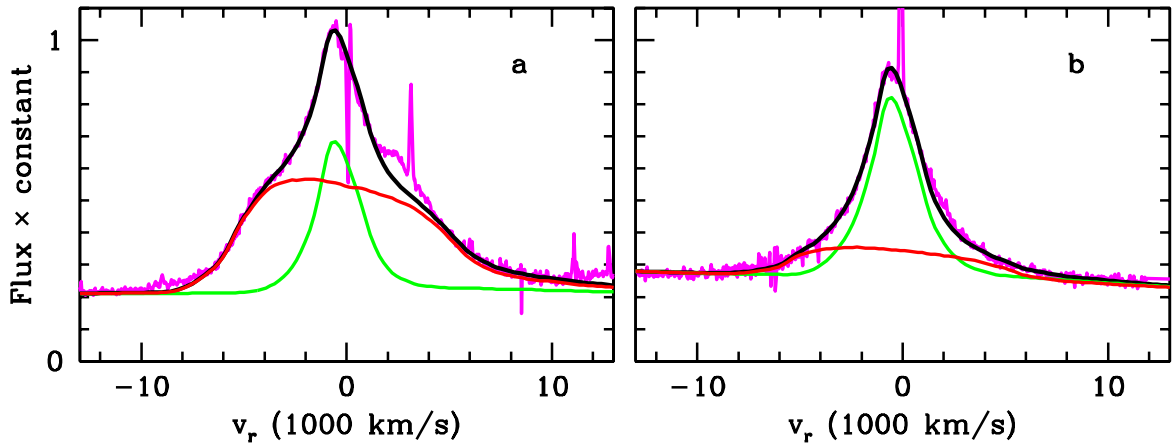}
    \caption{
    The He\,I 1.083 $\mu$m (panel {\bf a}) and Pa$\beta$ (panel {\bf b}) in SN~2010jl on day 178 
    ({\em magenta}) compared to models. The excess in the red wing of the He\,I 1.083 $\mu$m
    is due to Pa$\gamma$. Shown also contributions of the intermediate component ({\it green} line) and combined broad components of CDS and unshocked ejecta ({\it red}).    
    }
    \label{fig:he}
\end{figure*}

%
\begin{table}
\centering 
\caption[]{Parameters of demonstration models.}
\label{tab:hademo}
\begin{tabular}{l|c|c|c|c|c|c}
\hline
Model& $\tau_{cds}$ & $\tau_{sn}$  & $\omega_{sn}$ & $R_2$  & $f_{cds}$ & $f_{sn}$\\
\hline
 mod0  &  0.13   &  1    &  0.1   &   1.1 & 0.05 & 0.08\\
 mod1  &  0      &  0    &  0.1   &   1.1 & 0.05 & 0.08\\
 mod2  &  0      &  1    &  0.1   &   1.1 & 0.05 & 0.08  \\
 mod3  &  0.13   &  0    &  0.1   &   1.1 & 0.05 & 0.08 \\
 mod4  &  1.13   &  1    &  1     &   1.1 & 0.05 & 0.08\\
 mod5  &  0.13   &  1    &  0.1   &   1.2 & 0.05 & 0.08 \\
 mod6  &  0.13   &  1    &  0.1   &   1.1  & 0.2 & 0.2 \\
\hline
\end{tabular}
\label{tab:demo}
\end{table}

\subsection{Line profile}
\label{sec:prof} 

\subsubsection{Model}
\label{sec:mod}

A minimal model that captures major structure elements involved in the line profile formation 
of SN~2010jl includes three components: (i) freely expanding SN ejecta  
 ($v = r/t$ , $r < R_0 = 1$); (ii)  perturbed CDS ($R_0 < r < R_1 = 1.1$) with the average velocity $v_{cds}$ 
 and a random component in the range of $\delta v \sim 0.1v$ \citep[e.g.,][]{BloEllis2001}; 
(iii) line-emitting fragments of shocked CS clouds ($R_0 < r < R_2$) 
macroscopically mixed with the absorbing CDS material. This component is identified with the 
line intermediate component, while former two are responsible for the broad component. 
The narrow component originated in the preshock CSM is not considered in our model.
I neglect the hot thin layer 
in the reverse shock and set the radius of the unshocked SN ejecta equal to the CDS radius $R_0$. 

In line with arguments of Section \ref{sec:preview} the emissivity velocity distribution for 
the shocked CS clouds responsible for the intermediate component is approximated by the function
\begin{equation}
g(v) = 
 \left\{\begin{array}{ccc}
  \dfrac{v-v_{min}}{v_c-v_{min}}\,,  & \quad \textrm{if $v_{min}<v<v_c$} \\
  \left[\dfrac{v_{max}-v}{v_{max}-v_c}\right]^q\,,   & \quad \textrm{if $v_c<v<v_{max}$} \\
    0\,, & \quad \textrm{otherwise,}
    \end{array}\right.
\label{eq:vpdf}
\end{equation}
where the index $q$ is a free parameter close to unity. The function $g(v)$ has a maximum at $ v = v_c$ 
with the linear rise in the range $v_{c,min} < v < v_c$ that makes allowance for 
the fact that CS clouds are not identical; we adopt $v_{min} = 0.9v_c$. 
Note that in the case of $v_{min} = v_c$ the line profile 
would be flat in the range of radial velocities $\pm v_c$ which is not seen in observed profiles.
I preliminary explored two radial distribution of the emissivity of shocked clouds: (i) uniform 
in the range $R_0 < r < R_2$ (i.e., no radius/velocity correlation) and 
(ii) linear radius/velocity correlation, viz. $r=R_0 + (R_2-R_0)(v-v_{min})/(v_{max}-v_{min})$. 
The resulting profiles 
in these cases are indistiguishable, except for the late spectrum on day 804 when 
the correlation produces slightly better fit. This choice therefore is applied for all the epochs.

The broad component is emitted by the CDS and unshocked ejecta with relative fractions 
$f_{cds}$ and $f_{sn}$, so the intermediate component fraction is $1 - f_{cds} - f_{sn}$.
The CDS emissivity is distributed homogeneously in the velocity 
randomly distributed in the range of $v_{cds} < v < 1.1v_{cds}$.
The emissivity distribution in the SN ejecta is described by the "broken" power law 
$j(v) = (v/v_b)^{k_1}/[1 + (v/v_b)^{k_2}]$ with the velocity turnover $v_b$ and indices 
$k_1 \approx 2$ and $k_2 \approx 9$
inferred from broad component of the He\,I 1.083 $\mu$m line. The turnover of the $j(v)$ function
is expected because the ionizing radiation from the shocks is strongly absorbed in outer layers 
of ejecta.

The optical line emitted by ejecta can be scattered by electrons and dust (if any) and absorbed by the hydrogen 
and dust. The distribution of absorbers in SN ejecta and CDS is assumed to be homogeneous 
along the radius. 
The absorption is specified by the optical depth of the CDS  and ejecta ($\tau_{cds}$ and $\tau_{sn}$), 
and albedo ($\omega$), viz. the ratio of the scattering to extinction coefficient. To reduce the number of free 
parameters the albedo is set to be 
the same in the CDS and ejecta, which may not be the case. However this assumption does not affect significantly 
results because, 
as we will see, the ejecta dominates in the observable scattering effect.
In search for the optimal optical depth and albedo I ignore a possible lumpiness of the dust 
distribution. However, since effects of clumpy dust distribution will be discussed below, it 
is instructive to make some remarks concerning this issue. 

The effective optical depth in the clumpy medium is introduced \citet{Ambar1947} as
$\tau_{eff} = \tau_{oc}\langle 1 - \exp{(-\tau)}\rangle$,  
where $\tau_{oc}$ is the number of clouds along the ray, or the occultation optical depth, 
$\tau$ is the random cloud optical depth, and the averaging should be performed over cloud cross section 
and cloud ensemble. 
In the toy model of homogeneous spherical clouds with of the radius $b$ and the optical radius $\tau =kb$ 
homogeneously distributed in a sphere of a radius $R$
the average is calculated analytically \citep{Utrobin2015} and reads $\langle...\rangle = (4\tau/3)p(\tau)$, 
where $p(\tau)$ is the escape probability of photon emitted in the homogeneous sphere of the 
optical radius $\tau$ \citep[see][]{Osterbrock1989}. In the limit of $\tau \ll 1$ 
retaining only linear terms we get $\tau_{eff} = (4/3)\tau\tau_{oc}$ that can be easily transformed 
into $\tau_{eff} = fkR$, where $f$ is the filling factor of clouds. 
In the oposite case of $\tau \gg 1$ we obtain $\tau_{eff} = \tau_{oc}$ in line with the 
euristic definition of the optical depth for the ensemble of clouds treated as macroatoms. 

The optical depth recovered in the homogeneous model in clumpy case should be 
identified with the effective optical depth $\tau_{eff}$ and in the case of optically thick 
clouds this will be simply $\tau_{oc}$.
The albedo of the homogeneous model ($\omega$) 
for the clumpy medium should be associated with the albedo of a dusty cloud, which is expected to be 
lower than single scattering albedo of a grain ($\omega_d$).
The relation between $\omega_d$ and $\omega$ can be  
illustrated by Monte Carlo simulations for the spherical dusty cloud and spherical phase function.
In the limit of a large 
cloud optical radius ($\tau > 10$) for the set of the single scattering albedo (0.1, 0.3, 0.6, 0.9) the  
asymptotic values of $\omega$ turn out to be 0.023, 0.08, 0.2, and  0.5, respectively. 
If, e.g.,  one finds in the homogeneous model $\omega = 0.1$, then in the clumpy case with 
the large optical depth of the dusty cloud one expects the single scattering albedo to be  
$\omega_d \approx 0.35$. 

The apparent drawback of the outlined model is a large number of free parameters even in a
homogeneous case. The list includes $R_2$, $v_{sn}$, $v_{cds}$, $v_c$, $v_{max}$, $q$, $\tau_{cds}$, 
$\tau_{sn}$, $\omega$, $f_{cds}$ and $f_{sn}$, among which four 
 parameters $\tau_{cds}$, $\tau_{sn}$, $\omega$, and $v_{sn}$ are of prime interest for us.
The rest of parameters are constrained with different precisions. 
Particularly, at late time the contribution of the CDS is poorly constrained, so neither 
$f_{cds}$ and $v_{cds}$ can reliably inferred. Fortunately this does not affect the determination 
of the principal four parameters.
Although an effect of each parameter can be controlled, the problem of the unique 
set of parameters that defines the best fit model in multi-dimensional parameter space 
would be unproductive to solve rigorously, because the model is rather crude and ignores 
a possible deviation from the spherical symmetry that is plausible given a 
significant intrinsic polarization found in type IIn SN~1997eg \citep{Hoffman2008}, 
in some aspects reminding SN~2010jl. 

\subsubsection{Results}

In a search for the optimal model I explored an extended volume of parameter space.
Fortunately effects of most parameters are rather obvious, although in some cases the 
effect is less trivial. To provide insight into the role of principal paprameters I present 
a set of simulated Lorentz-like profiles (Table \ref{tab:demo}, Fig. \ref{fig:demo}). 
Each panel shows a template model (mod0) that corresponds to the H$\alpha$ model on day 510 
and a model with the altered parameter value.  The template model 
is dominated by the emission of shocked CS clouds (intermediate component); ejecta and CDS 
contribute together 10\% (panel {\bf a}). 
The unabsorbed line with zero optical depth for both CDS and SN ejecta (panel {\bf a}) 
shows symmetriic profile with roundish top that is due to the low velocity increasing 
part of the velocity spectrum $g(v)$ of shocked CS clouds (cf. Section \ref{sec:mod}).
The case of a zero CDS absorption (panel {\bf b}) demonstrates that the ejecta absorption 
is not sufficient to produce required blueshift of the line maximum. On the other hand, 
the absence of the ejecta absorption (panel {\bf c}) shows that while the CDS absorption
is able to describe the blueshift of the maximum, the profile in the red is unlike what is required. 
Note that the dip in the red part is a limb effect for the line arising in a narrow shell ($\Delta R/R < 1$)
with the continuum 
absorption/scattering \citep[cf.][]{Chugai1991}. The case of $\omega = 1$ corresponding to 
the Thomson scattering with the thermal broadening taken inro account results in the pronounced red 
wing (panel {\bf d}), a natural outcome of the photon scattering in the 
expanding envelope \citep{Auer1972}. This effect is also present in the case when line 
photons are scattered off a dust with albedo $\omega \sim 0.5$ in SN ejecta \citep{Lucy1989,Bevan2016}. 
The need for the macroscopic mixing between the layer of shocked CS clouds and the dusty CDS material 
is demonstrated by model with the outward shift of the line-emitting layer by $\Delta r = 0.1R_0$          
(panel {\bf e}): the model shows lower maximum blueshift. Interestingly, the profile in the model mod5 
almost exactly repeats the profile in the model mod2. This is not unexpected since in 
both cases line-emitting shell lies above the absorbing shell. This similarity however cannot be considered 
as the parameter degeneracy, because in the model with correct blueshift these similar cases cannot appear.
The effect of large contribution of the 
broad components related to CDS and SN ejecta 
indicates how the broad components can be constrained (panel {\bf f}).
The value of the index $q$ affects the intermediate component in a sense that $q \gg 1$ makes 
profile narrower and more gaussian-like, while $q \ll 1$ makes profile broader and closer to the 
triangular shape. 

We now consider an interesting case of the He\,I 1.083 $\mu$m line.
At the early stage ($t < 400$ d)  He\,I~1.083 $\mu$m and He\,I~2.06 $\mu$m lines 
show a pronounced broad component unlike smooth hydrogen line profiles \citep{Borish2015}. 
A reasonable suggestion is that this component is caused by 
the emission of the SN ejecta \citep{Borish2015}. 
Figure \ref{fig:he} demonstrates satisfactory fits He\,I 1.083 ${\mu}$m and Pa$\beta$ on day 178
in the models with the same parameters 
except for the ejecta emission fraction, $f_{sn} = 0.68$ for He\,I 1.083 ${\mu}$m 
compared to 0.3 for for Pa$\beta$, and the index, $q = 1.5$ for He\,I 1.083 ${\mu}$m compared to 
$q = 0.9$ for Pa$\beta$.  The latter difference
presumably reflects different excitation conditions of hydrogen and helium  
in the intermediate component. 
Both helium and hydrogen line shows blueshift that is 
expressed in the inferred values of optical depths $\tau_{cds} = 0.07$, $\tau_{sn} = 0.58$, and 
albedo $\omega = 0.6$. Since at this epoch the dust formation in the CDS and SN ejecta is 
unlikely, the extiction is probaly related to the hydrogen photoabsorption and Thomson scattering. 
It should be emphasised that the blueshift of the broad component of He\,I 1.083 ${\mu}$m is produced 
entirely by the absorption in the unshocked ejecta.
It is not clear whether the different fraction of the broad component for helium and hydrogen 
lines is related to the different helium abundance in SN ejecta and CSM, or different 
excitation conditions in the line-emitting regions. Discussion of this issue would
require highly complicated modelling, which is currently beyond reach.
The recovered parameters of the He\,I ejecta component, viz. 
$v_b = 5500$\,\kms, $k_1 = 2$, and $k_2 = 9$  (cf. Equation \ref{eq:vpdf}) 
are adopted henceforth for hydrogen lines as well because in hydrogen lines the ejecta component 
parameters cannot be reliably constrained. 

\begin{table*}
\centering
\caption[]{Parameters of H$\alpha$ models.}
\label{tab:hamod}
\begin{tabular}{l|c|c|c|c|c|c|c|c|c|c|c}
\hline
Model & $v_{sn}$  & $v_{cds}$ &  $v_{c}$ &  $v_{max}$ & $q$ & $R_2$ & 
$\tau_{cds}$ & $\tau_{sn}$ & $\omega$ & $f_{cds}$ & $f_{sn}$  \\
\hline
& \multicolumn{4}{c} {km\,s$^{-1}$} & & & & & & &  \\
\hline
 m172   &  8500   & 5600   &  900 & 5300  & 0.9    &  1.13 & 
  0.07  & 0.58    & 0.6    & 0.05 & 0.3  \\
 m400   &  6100   & (4700)  & 1100 & 4400  & 0.6   & 1.1 & 
  0.14  & 1.15     & 0.2    & 0.05 & 0.12  \\ 
 m510   &  5700   & (4200)   & 850 & 3500   & 0.4   & 1.1 & 
  0.12  & 0.99     & 0.08  & 0.05 & 0.08  \\  
 m804   &  (5000)  &  (3200)   & 850 & 3000   & 1.2   & 1.1 & 
  0.18   & 0.96       & 0.1    & (0.02) & (0.02)  \\  
 \hline
 \end{tabular}
 \label{tab:param}
\end{table*}

\begin{figure*}
	\includegraphics[trim=60 170 0 -20,width=0.95\textwidth]{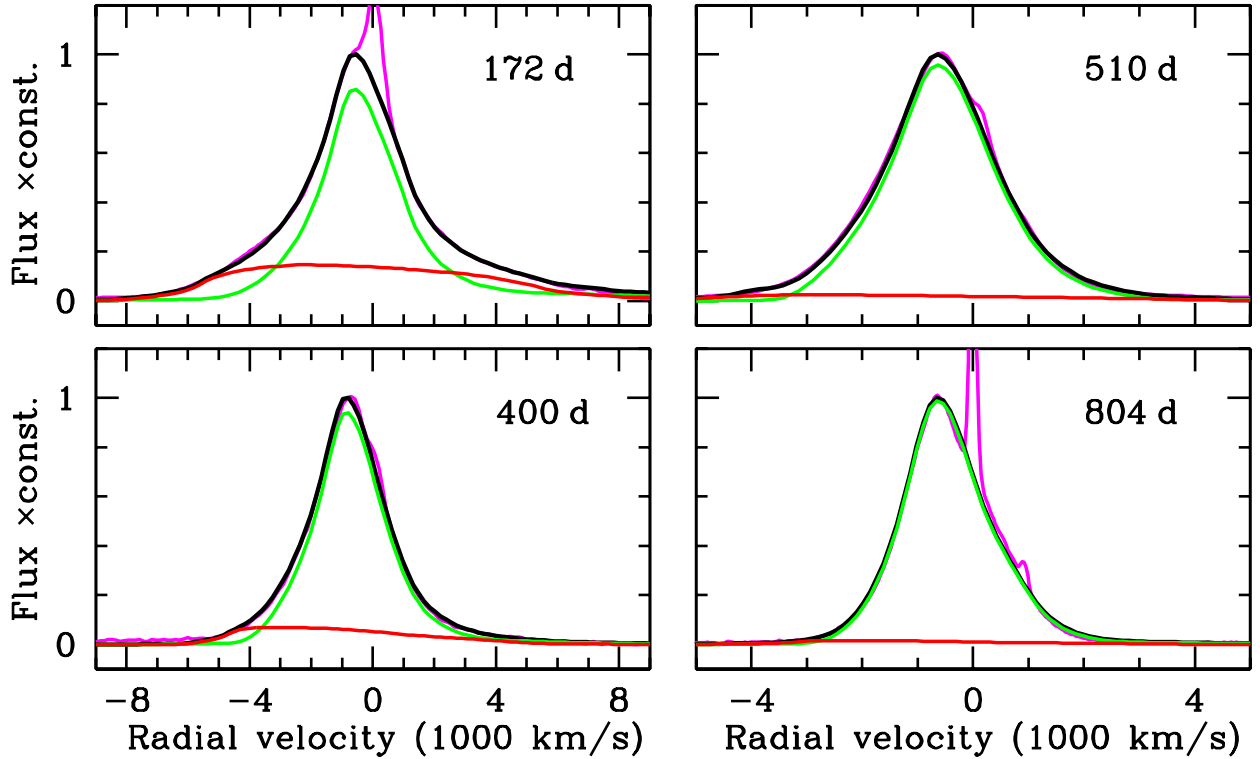}
    \caption{
    H$\alpha$ line in SN~2010jl ({\em magenta}) compared to the model profile at 
    four epochs ({\it black} line). The blushifted profiles are well reproduced for adopted parameters 
    (Table \ref{tab:param}). The excess on day 172 around zero velocity is the narrow component, 
    while on day 804 the excess is due to superimposed H\,II region.
    Shown are also contributions of the dominant intermediate component ({\em green})
     and combined broad components 
    of CDS and unshocked ejecta ({\it red}). 
     }
    \label{fig:hamod}
\end{figure*}

The H$\alpha$ line is modelled for spectra taken on 2011 March 31 (172 days), 
2011 Nov. 13 (400 days), 2012 Feb. 28 (510 days), and 2012 Dec. 21 (804 days). The data  
are retrieved 
from  the Weizmann supernova data repository {\em http://wiserep.weizmann.ac.il} 
\citep{Yaron2012}. Models overplotted on the observed spectra 
are displayed in Fig. \ref{fig:hamod} with parameters given in the Table \ref{tab:hamod}.
Columns one by one contain: the model name (the number is the age in days), terminal velocity 
of unshocked SN ejecta ($v_{sn}$), CDS velocity ($v_{cds}$), 
velocity of CS cloud shock ($v_{c}$), maximal velocity of the CS shocked cloud material ($v_{max}$),
index $q$ of the emissivity of shocked CS clouds, 
outer radius of the line-emitting layer of shocked CS clouds ($R_2$);
continuum optical depths of the CDS ($\tau_{cds}$) and unshocked SN ejecta ($\tau_{sn}$);
albedo ($\omega$) and the relative contribution of the CDS ($f_{cds}$) and unshocked SN ($f_{sn}$) into 
the H$\alpha$ luminosity. The values in parentheses have large uncertainties.
For all the considered epochs the CDS contribution is dominant. 
The contribution of the broad components both ejecta and the CDS is large only in the spectrum 
on day 172 and gets small with the age. 

Remarkably, the CDS optical depth at the late stage (510 and 804 d), when presumably the dust have 
been formed, is rather small (0.13 and 0.15 respectively), whereas SN ejecta turns out essentially 
more opaque  ($\tau_{sn} = 1$). 
This indicates that apart from the CDS the dust also forms in the unshocked ejecta and possibly in 
larger amount. Note that at late time the lower albedo is preferred. This fact is consistent with 
the dust to be the main source of opacity, since the hydrogen density and hydrogen excitation 
in the SN ejecta decreses with time and therefore the hydrogen absorption and Thomson scattering 
gets inefficient.

The estimated parameter errors that reflect precision of model fits lie in the range 10-20\%
except for those inicated by parentheses in Table \ref{tab:param}. The errors however 
reflect only the adopted model. In fact, the radial distribution of 
absorbers and emitters in ejecta and velocity and radial distributions of shocked CS clouds might  
be more complicated; moreover, the anisotropy is not rulled out. All these can result in some
change of values of four principal parameters ($\tau_{cds}$, $\tau_{sn}$, $\omega$, and $v_{sn}$).
With this remark, the proposed spectral model should be considered as a reasonable    
possibility in the absence of alternative models.

\section{CS interaction and dusty shell}
\label{sec:dmod} 

The interaction of the SN ejecta with the CSM is analysed using the hydrodynamic
model in a thin shell approximation \citep{Giuliani1982,Chevalier1982a}. 
The central to the model is the equation of motion of the swept up thin shell driven by the 
dynamic pressure of SN ejecta $\rho_{sn}(R_s/t - v_s)^2$ and dynamic pressure of CSM 
$\rho_{cs}(v_s - v_w)^2$, where $R_s$ and $v_s$ are the radius and velocity of the thin shell,
$v_w$ is the wind velocity taken to be 100\,\kms\ \citep{Fransson2014}. 
The model has been described in details earlier \citep{Chugai2001}, here we recap only key points.
We do not distinguish between the swept up shells formed in the reverse and forward 
shocks: both are presumably merged into a single shell.  
The optical luminosity of the reverse and forward shocks at the age $t$ are calculated as 
the shock kinetic luminosities $0.5\rho_{sn}(R_s/t - v_s)^3$ and $0.5\rho_{cs}(v_s - v_w)^3$ 
multiplied by the radiation efficiency $\eta = t/(t+t_c)$ of a relevant shock, where 
$t_c$ is the cooling time. 
The radiation escape from both shocks is assumed to be instant, i.e., weak diffusion effects are ignored.
The model does not include an additional luminosity related to the internal energy of an explosion. 

The cooling time is calculated neglecting the lumpiness of the CSM and adopting postshock density to be 
equal to four-fold preshock density characteristic of a strong adiabatic shock with the adiabatic 
index $\gamma=5/3$. 
This approximation describes the energetically significant initial stage of the radiative shock cooling.
The reverse shock in the considered case is always strongly radiative, while the forward 
shock is partially radiative with $\eta \sim 0.5$ between days 200 and 900.
The lumpiness of the CSM could modify the cooling time of the forward shock. The effect however 
is not easy to estimate 
without detailed 3D-modelling. At first glance the presence of the dense shocked CS clouds should accelerate 
the radiative cooling. However, in a collision of forward shock 
rarefied flow and dense CS cloud most of the dissipated kinetic luminosity is deposited into the rarefied flow 
which results in the slowdown of the cooling process thus compensating effect of the dense 
shocked clouds.
We therefore do not expect a significant modification of the bolometric light curve.
As to the CDS dynamics, it is not affected by the cooling time modification.

The adopted density distribution of SN ejecta is \mbox{$\rho = \rho_0/(1 + (v/v_0)^8)$} with 
$\rho_0$ and $v_0$ determined by the ejecta mass and kinetic energy. The 
result is not very sensitive to the ejecta mass, nevertheless we consider two cases of ejecta mass:
$M_{sn} = 8\,M_{\odot}$,  
the value used earlier for SN~2010jl \citep{Chandra2015}, and $M_{sn} = 30\,M_{\odot}$.
The remaining parameters are fixed by fitting 
to the bolometric light curve, CDS, ejecta velocities (Table \ref{tab:param}), and the CSM column density. 
The CSM density is set by a broken power low $\rho \propto r^{-s}$ with $s = 0$ in the inner region 
$r < 3.1\times10^{15}$ cm, $s \sim 1.6...~2$ in the intermediate zone $r < 4\times10^{16}$ cm, and 
steep drop ($s = 11$) in the outer region.

\begin{figure}
	\includegraphics[trim=70 100 0 -10,width=\columnwidth]{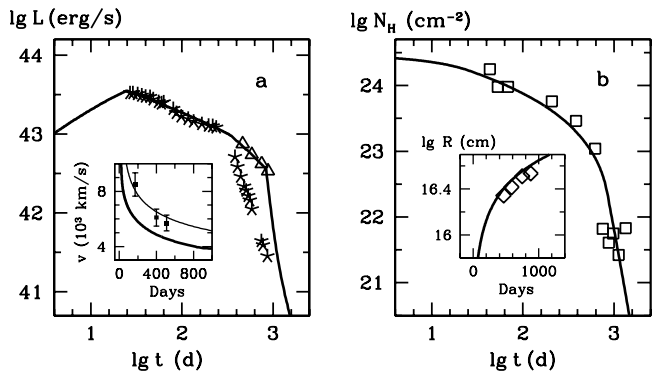}
    \caption{
    The SN/CSM interaction model for the ejecta mass of $8\,M_{\odot}$ with observational data. 
    Panel {\bf a} shows the bolometric light curve compared to observational 
    data from \citet{Fransson2014} ({\em asterisks}; IR luminosity summed with the optical 
    is shown by {\em triangles}). 
    Inset shows the boundary velocity of the unshocked ejecta ({\it thin} line) compared to data 
    recovered from the H$\alpha$ and the CDS velocity  ({\it thick} line)
    Panel {\bf b} shows the evolution of the CSM column density ahead of the forward 
    shock compared to data ({\em squares}) \citep{Chandra2015}. Inset shows the CDS radius 
    compared to  the radius of the IR-emitting shell ({\em diamonds}) \citep{Fransson2014}.
    }
    \label{fig:dyn1}
\end{figure}

\begin{figure}
	\includegraphics[trim=70 100 0 -10,width=\columnwidth]{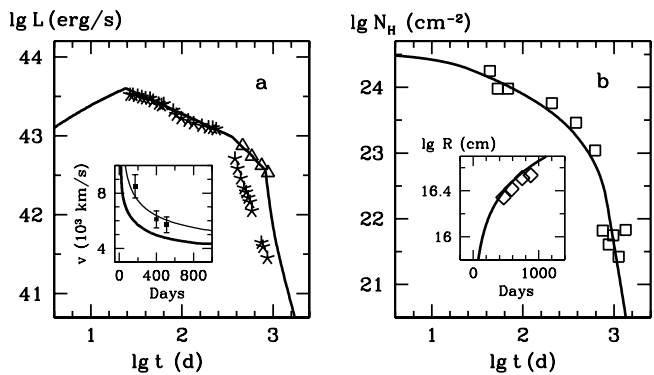}
    \caption{
    The same as Fig. \ref{fig:dyn1} but for the ejecta mass of $30\,M_{\odot}$
    }
    \label{fig:dyn2}
\end{figure}

 The model with the ejecta mass of $8\,M_{\odot}$ and  kinetic energy of $E = 3.9\times10^{51}$ erg 
describes the observational bolometric luminosity, ejecta velocities, and  
CSM column density (Fig. \ref{fig:dyn1}).
Note that the observational bolometric light curve \citep{Fransson2014} is modified by the inclusion of
the late  ($t > 460$ d) infrared luminosity reported by \citet{Fransson2014}. 
The total mass of the CSM turns out to be $M_{cs} = 3.9~M_{\odot}$.
The model with $M_{sn} = 30\,M_{\odot}$ and $E = 6.9\times10^{51}$ erg (Fig. \ref{fig:dyn2})
describes data with the same CSM mass $M_{cs} = 3.8~M_{\odot}$. 
Both models fit the data well and neither option is preferred.
The main outcome of the modelling is the fact that in both models the CDS/SN radius at 
late time $t > 460$ d (Fig. \ref{fig:dyn1},~\ref{fig:dyn2}) 
coincides with the radius of the dusty shell reported by \citet{Fransson2014}. 
This strongly suggests that the dust of CDS and/or unshocked ejecta is responsible for the 
late IR luminosity.

\section{Discussion}
\label{sec:disc}

The aim of the paper has been to test a conjecture that in the late time SN~2010jl 
the dust forms in the CDS. To this end I propose a model explaining the Lorentz-like 
line profile as an outcome of hydrogen line emission from from shocked CS clumps in the 
forward shock. It is found that the macroscopic mixing of CS cloud 
fragments with the absorbing CDS and a significant absorption in the unshocked 
ejecta accounts for the observed blueshift.
Noteworthy that the blueshift of hydrogen and He\,1.083 $\mu$m lines at the early stage (< 200 d) 
may not be related with the dust 
absorption at all; the opacity is likely maintained by the excited hydrogen. The remarkable 
result suggested by the line profile modelling is that the absorption in the CDS is not sufficient: 
a significant absorption in the unshocked SN ejecta is needed to account for the whole blueshifted 
line profile. Moreover, the required optical depth 
of ejecta is 5-7 times larger compared  to the CDS at the epoch 500-800 days. This indicates that 
the dust should form not only in the CDS but in the unshocked ejecta as well.

I found that the SN/CSM interaction model tuned by the fit to 
the bolometric light curve, ejecta velocities recovered from 
H$\alpha$, and the CSM column density \citep{Chandra2015} predicts the outer SN radius 
at late time ($t > 460$\,d) consistent with the radius of the dusty IR-emitting shell 
recovered by \citet{Fransson2014}. The latter fact combined with the line blueshift  
analysis strongly suggests that at late time (> 460 d) the dust indeed forms both in the CDS and unshocked 
ejecta. Thus the conjecture of the dust formation in the CDS
\citep{Maeda2013,Gall2014} is only partially true: a significant 
amount of the dust should form in the unshocked ejecta. 
Moreover, the low  covering factor of the CDS indicated 
by the small CDS optical depth $\sim 0.2$ at 510\,d and 804\,d suggests that the 
effective radius of the IR-emitting CDS photosphere turns out to be twice as small compared to 
the reported radius of the dusty shell \citep {Fransson2014}. This means that 
the late IR emission is likely primarily related to the more opaque dusty SN ejecta.

The radius of the late dust-emitting shell have been obtained \citep{Fransson2014} using black 
body approximation, which suggests that the IR-emitting ejecta should be optically thick 
in the near IR band, e.g., at 3\,$\mu$m the optical depth must be $ > 1$. The direct consequence of 
that is the large optical depth at the H$\alpha$ ($\tau > 10$) assuming the absorption 
efficiency $Q_a \propto \lambda^{-1.5}$. This optical depth   
is by an order of magnitude larger than the value inferred from the H$\alpha$ profile between days 500 and 
800 (Table \ref{tab:param}). The tension can be resolved assuming that the dust in ejecta 
is locked in opaque clumps, likewise in the model proposed for the dusty zone of 
SN~1987A ejecta \citep{Lucy1991}. Remarkably, the dust distribution in opically thick clumps for SN~2010jl 
has been already proposed by \citet{Maeda2013}. 
In the case of clumpy dust distribution in unshocked SN ejecta the optical depth recovered from the 
blueshift assuming homogeneous model 
should be identified with the occultation optical depth $\tau_{oc}$, i.e., the number of clouds 
intersecting by the ejecta radius $R$. 

One can estimate the amount of dust in SN ejecta adopting a simple model 
of opaque spherical dusty blobs of a radius $b$ homogeneously distributed in SN ejecta. 
The occultation optical depth is then  
$\tau_{oc} = \pi b^2n_cR = (3/4)(b/R)^2N_c$, 
where $n_c = 3N_c/(4\pi R^3)$ is the number density of clouds and $N_c$ is the total number of 
clouds. The optical radius of the individual cloud 
is $\tau_c = n_g\langle \sigma_a \rangle b$, 
where $n_g$ is number density of grains, and $\langle \sigma_a \rangle$ is the absorption cross section of a 
grain averaged over the grain ensemble assuming grain radii distribution 
$dn/da \propto a^{-3.5}$ ($a_{min}=3\times10^{-7}$\,cm, 
$a_{max}=3\times10^{-5}$\,cm) and using absorption efficiency of the carbon and silicate dust according to
\citet{Draine1984}. 
The total amount of the dust in SN ejecta is $M_d =(4\pi/3)b^3n_c\langle m_g\rangle N_c$, where 
$\langle m_g\rangle$  is the grain mass avaraged over the grain ensemble assuming the grain density 3\,g\,cm$^{-3}$.
Expressing $N_c$ via $\tau_{oc}$ and $b$ via $\tau_c$ the dust mass thus reads
\begin{equation}
M_d = (16\pi/9)R^2\tau_c\tau_{oc}\langle m_g\rangle/\langle \sigma_a \rangle\,.
\label{eq:mdust}
\end{equation}
Inserting $\tau_{oc} = 1$, the value derived from H$\alpha$ on day 804, and the ejecta radius 
$R = 3.8\times10^{16}$ cm 
obtained in the interaction model we infer from equation (\ref{eq:mdust}) the dust mass of
$7.7\times10^{-4}\tau_b\,M_{\odot}$ for the graphite and $3.9\times10^{-3}\tau_b\,M_{\odot}$ for silicate dust.
For optically thick blobs with $\tau_b = 2$ at 3 $\mu$m the amount of dust in ejecta is 
$8\times10^{-3}\,M_{\odot}$ for the silicate and $1.5\times10^{-3}\,M_{\odot}$ for the graphite dust. 
The conservative estimate of the dust amount is therefore $> 10^{-3}\,M_{\odot}$.

In the IR data reported by \citet{Fransson2014} a signature of the hot dust becomes apparent only
on day 465. In this regard, the blueshift at the early epoch (e.g. 172 d) is unlikely related to the dust, 
rather it is due to the hydrogen opacity. To illustrate the possibility, consider the 
situation on day 172. Assuming the excitation temperature equal to the spectrophotometric 
estimate of 7500\,K \citep[cf.][]{Fransson2014} and the CDS mass of $\approx 2\,M_{\odot}$, according to the 
interaction model (Section \ref{sec:dmod}), the optical depth due to the hydrogen continuum absorption 
at 6500 \AA\ assuming Saha-Boltzmann population distribution turns out to be $\sim 0.5$,  
sufficient to account for the CDS optical depth in the 
model m172. This consideration emphasises the point that the line blueshift in SNe~IIn at early epoch 
cannot be a reliable argument in favour of the dust formation in CDS and/or ejecta unless supported 
by the blueshift emergence on the time scale $\ll t$.

The observed phenomena in late SN~2010jl, of course, are more complicated than  
our simplified consideration suggests. Particularly, we did not address the ratio of escaping 
optical-to-IR fluxes, implicitly assuming that 
the late IR luminosity is a result of the radiation cascade of the XUV radiation 
of the forward and reverse shocks. 
Meanwhile, at late time the optical-to-IR ratio is surprisingly small: 
around day 800 the difference is a factor of ten \citep{Fransson2014}. 
That low ratio cannot be explained in the picture in which the primary optical emission is comparable to 
the IR radiation and only the dust absorption is responsible for the low escaping optical luminosity. 
Indeed, despite the absorption, almost half of the emitted optical radiation could
escape because of the small CDS optical depth ($\sim 0.2$).
The low optical-to-IR ratio at late time might be related to the conversion of the  
 XUV radiation in the cool gas directly into the 
IR radiation avoiding significant emission of the optical radiation. 
This conjecture could be verified by the detection of strong 
IR features related to molecular species in supernovae of the similar category at the similar epoch 
(1.5-2 yr). An alternative  possibility is that
at late time the forward shock becomes adiabatic which accounts for the low optical luminosity 
while the high IR luminosity originates from the collisional heating of the newly formed dust mixed with 
the hot postshock gas.

Another puzzling issue is the weakness of the H$\alpha$ broad boxy component 
related to the CDS material. A plausible answer can be seen from the following arguments. 
The strong upper limit for the H$\alpha$ luminosity of a thin dense spherical shell with the radius $R$ in the 
case of very large line optical depth ($\tau > 10^4$) is 
\begin{equation}
L(\mbox{H}\alpha) = 4\pi^2R^2B_{\nu}(T)x_e\Delta \nu_D\,,
\label{eq:halum}
\end{equation}
where $B_{\nu}(T)$ is the black body intensity, $\Delta \nu_D$ is the thermal Doppler width 
for hydrogen, 
$x_e = (a\tau/\sqrt{\pi})^{1/3}$ is the local line width in units of the Doppler width 
in the case of the very opaque layer \citep{Adams1975}, 
$a =3.3\times10^{-3}$ is the Voigt parameter for the H$\alpha$ for $T \approx 7500$\,K. 
For SN~2010jl on day 200 the shell radius is $R = 1.4\times10^{16}$ cm, 
and the CDS mass is $\approx 2\,M_{\odot}$ (according to the interaction model in Section \ref{sec:dmod}). 
Adopting the excitation 
temperature of 7500\,K \citep{Fransson2014}, and assuming that the hydrogen is mostly 
neutral (which majorizes line width $x_e$) for solar hydrogen abundance one gets 
$L(\mbox{H}\alpha) = 0.9\times10^{41}$ erg\,s$^{-1}$. 
Note that assumptions made above strongly overestimate the luminosity. Even in that case 
the obtained value is a factor of ten  
lower compared to the observed H$\alpha$ luminosity of $\approx 10^{42}$ erg\,$s^{-1}$ 
\citep{Fransson2014} at that stage.  
If the CDS is not strongly disturbed, so that the area ratio for the cumulative surface 
$S$ of the line-emitting CDS material $A = S/4\pi R^2 \sim 1$, then the broad boxy H$\alpha$ 
related to the CDS turns out relatively weak ($ < 10$\%) and therefore gets overwhelmed by
the broad wings of the ejecta component. 
The above argument have an interesting implication: the dominant contribution of 
the shocked CS clumps in the H$\alpha$ luminosity suggests that the area ratio of the 
cumulative surface of the line-emitting CS cloud fragments should be large, of the order of $A \sim 10$.

\section{Conclusion}

I propose a simple model that accounts for the blueshifted line profiles in spectra of 
SN~2010jl at late time ($\geq  500$ d). The key elements of this picture are 
(i) the shocked CS clouds in the forward shock, which are responsible for the dominant 
intermediate component of line profiles, (ii) the dusty cool dense shell with the small 
optical depth that accounts for the blueshift of the line maximum, and (iii) dusty unshocked 
SN ejecta that are responsible for the overall line blueshift. 
The modelling of the SN interaction with the CSM shows that the SN 
radius is consistent with the radius of optically thick dusty 
shell recovered from IR data. This taken together with the line 
profile analysis suggests that the dust formed in CDS and  
in unshocked SN ejecta is responsible for both the line blueshift and 
the IR emission. Most of the dust reside in optically thick clumps of 
the unshocked ejecta.

\section*{Acknowledgements}



\bibliographystyle{mnras}
\bibliography{sn2010jl.bib}

\begin{thebibliography}{}
\makeatletter
\relax
\def\mn@urlcharsother{\let\do\@makeother \do\$\do\&\do\#\do\^\do\_\do\%\do\~}
\def\mn@doi{\begingroup\mn@urlcharsother \@ifnextchar [ {\mn@doi@}
  {\mn@doi@[]}}
\def\mn@doi@[#1]#2{\def\@tempa{#1}\ifx\@tempa\@empty \href
  {http://dx.doi.org/#2} {doi:#2}\else \href {http://dx.doi.org/#2} {#1}\fi
  \endgroup}
\def\mn@eprint#1#2{\mn@eprint@#1:#2::\@nil}
\def\mn@eprint@arXiv#1{\href {http://arxiv.org/abs/#1} {{\tt arXiv:#1}}}
\def\mn@eprint@dblp#1{\href {http://dblp.uni-trier.de/rec/bibtex/#1.xml}
  {dblp:#1}}
\def\mn@eprint@#1:#2:#3:#4\@nil{\def\@tempa {#1}\def\@tempb {#2}\def\@tempc
  {#3}\ifx \@tempc \@empty \let \@tempc \@tempb \let \@tempb \@tempa \fi \ifx
  \@tempb \@empty \def\@tempb {arXiv}\fi \@ifundefined
  {mn@eprint@\@tempb}{\@tempb:\@tempc}{\expandafter \expandafter \csname
  mn@eprint@\@tempb\endcsname \expandafter{\@tempc}}}

\bibitem[\protect\citeauthoryear{{Adams}}{{Adams}}{1975}]{Adams1975}
{Adams} T.~F.,  1975, \mn@doi [\apj] {10.1086/153891}, \href
  {http://adsabs.harvard.edu/abs/1975ApJ...201..350A} {201, 350}

\bibitem[\protect\citeauthoryear{{Ambarzumian}}{{Ambarzumian}}{1947}]{Ambar1947}
{Ambarzumian} V.~A.,  1947, Doklady Acad. Sci. Armenia SSR, 6, 105 ({\it in
  russian})

\bibitem[\protect\citeauthoryear{{Andrews} et~al.,}{{Andrews}
  et~al.}{2010}]{Andrews2010}
{Andrews} J.~E.,  et~al., 2010, \mn@doi [\apj] {10.1088/0004-637X/715/1/541},
  \href {http://adsabs.harvard.edu/abs/2010ApJ...715..541A} {715, 541}

\bibitem[\protect\citeauthoryear{{Andrews} et~al.,}{{Andrews}
  et~al.}{2011}]{Andrews2011}
{Andrews} J.~E.,  et~al., 2011, \mn@doi [\aj] {10.1088/0004-6256/142/2/45},
  \href {http://adsabs.harvard.edu/abs/2011AJ....142...45A} {142, 45}

\bibitem[\protect\citeauthoryear{{Auer} \& {van Blerkom}}{{Auer} \& {van
  Blerkom}}{1972}]{Auer1972}
{Auer} L.~H.,  {van Blerkom} D.,  1972, \mn@doi [\apj] {10.1086/151777}, \href
  {http://adsabs.harvard.edu/abs/1972ApJ...178..175A} {178, 175}

\bibitem[\protect\citeauthoryear{{Bevan} \& {Barlow}}{{Bevan} \&
  {Barlow}}{2016}]{Bevan2016}
{Bevan} A.,  {Barlow} M.~J.,  2016, \mn@doi [\mnras] {10.1093/mnras/stv2651},
  \href {http://adsabs.harvard.edu/abs/2016MNRAS.456.1269B} {456, 1269}

\bibitem[\protect\citeauthoryear{{Blondin} \& {Ellison}}{{Blondin} \&
  {Ellison}}{2001}]{BloEllis2001}
{Blondin} J.~M.,  {Ellison} D.~C.,  2001, \mn@doi [\apj] {10.1086/322499},
  \href {http://adsabs.harvard.edu/abs/2001ApJ...560..244B} {560, 244}

\bibitem[\protect\citeauthoryear{{Borish}, {Huang}, {Chevalier}, {Breslauer},
  {Kingery}  \& {Privon}}{{Borish} et~al.}{2015}]{Borish2015}
{Borish} H.~J.,  {Huang} C.,  {Chevalier} R.~A.,  {Breslauer} B.~M.,  {Kingery}
  A.~M.,   {Privon} G.~C.,  2015, \mn@doi [\apj] {10.1088/0004-637X/801/1/7},
  \href {http://adsabs.harvard.edu/abs/2015ApJ...801....7B} {801, 7}

\bibitem[\protect\citeauthoryear{{Chandra}, {Chevalier}, {Irwin}, {Chugai},
  {Fransson}  \& {Soderberg}}{{Chandra} et~al.}{2012}]{Chandra2012}
{Chandra} P.,  {Chevalier} R.~A.,  {Irwin} C.~M.,  {Chugai} N.,  {Fransson} C.,
    {Soderberg} A.~M.,  2012, \mn@doi [\apjl] {10.1088/2041-8205/750/1/L2},
  \href {http://adsabs.harvard.edu/abs/2012ApJ...750L...2C} {750, L2}

\bibitem[\protect\citeauthoryear{{Chandra}, {Chevalier}, {Chugai}, {Fransson}
  \& {Soderberg}}{{Chandra} et~al.}{2015}]{Chandra2015}
{Chandra} P.,  {Chevalier} R.~A.,  {Chugai} N.,  {Fransson} C.,   {Soderberg}
  A.~M.,  2015, \mn@doi [\apj] {10.1088/0004-637X/810/1/32}, \href
  {http://adsabs.harvard.edu/abs/2015ApJ...810...32C} {810, 32}

\bibitem[\protect\citeauthoryear{{Chevalier}}{{Chevalier}}{1982a}]{Chevalier1982b}
{Chevalier} R.~A.,  1982a, \mn@doi [\apj] {10.1086/160126}, \href
  {http://adsabs.harvard.edu/abs/1982ApJ...258..790C} {258, 790}

\bibitem[\protect\citeauthoryear{{Chevalier}}{{Chevalier}}{1982b}]{Chevalier1982a}
{Chevalier} R.~A.,  1982b, \mn@doi [\apj] {10.1086/160167}, \href
  {http://adsabs.harvard.edu/abs/1982ApJ...259..302C} {259, 302}

\bibitem[\protect\citeauthoryear{{Chevalier} \& {Blondin}}{{Chevalier} \&
  {Blondin}}{1995}]{CheBlo1995}
{Chevalier} R.,  {Blondin} J.~M.,  1995, \mn@doi [\apj] {10.1086/175606}, \href
  {http://adsabs.harvard.edu/abs/1995ApJ...444..312C} {444, 312}

\bibitem[\protect\citeauthoryear{{Chugai}}{{Chugai}}{1991}]{Chugai1991}
{Chugai} N.~N.,  1991, \mn@doi [\mnras] {10.1093/mnras/250.3.513}, \href
  {http://adsabs.harvard.edu/abs/1991MNRAS.250..513C} {250, 513}

\bibitem[\protect\citeauthoryear{{Chugai}}{{Chugai}}{2001}]{Chugai2001}
{Chugai} N.~N.,  2001, \mn@doi [\mnras] {10.1111/j.1365-8711.2001.04717.x},
  \href {http://adsabs.harvard.edu/abs/2001MNRAS.326.1448C} {326, 1448}

\bibitem[\protect\citeauthoryear{{Chugai}}{{Chugai}}{2009}]{Chugai2009}
{Chugai} N.~N.,  2009, \mn@doi [\mnras] {10.1111/j.1365-2966.2009.15506.x},
  \href {http://adsabs.harvard.edu/abs/2009MNRAS.400..866C} {400, 866}

\bibitem[\protect\citeauthoryear{{Chugai} \& {Danziger}}{{Chugai} \&
  {Danziger}}{1994}]{Chugai1994}
{Chugai} N.~N.,  {Danziger} I.~J.,  1994, \mn@doi [\mnras]
  {10.1093/mnras/268.1.173}, \href
  {http://adsabs.harvard.edu/abs/1994MNRAS.268..173C} {268, 173}

\bibitem[\protect\citeauthoryear{{Draine} \& {Lee}}{{Draine} \&
  {Lee}}{1984}]{Draine1984}
{Draine} B.~T.,  {Lee} H.~M.,  1984, \mn@doi [\apj] {10.1086/162480}, \href
  {http://adsabs.harvard.edu/abs/1984ApJ...285...89D} {285, 89}

\bibitem[\protect\citeauthoryear{{Dwek} et~al.,}{{Dwek}
  et~al.}{2017}]{Dwek2017}
{Dwek} E.,  et~al., 2017, \mn@doi [\apj] {10.3847/1538-4357/aa8665}, \href
  {http://adsabs.harvard.edu/abs/2017ApJ...847...91D} {847, 91}

\bibitem[\protect\citeauthoryear{{Elmhamdi} et~al.,}{{Elmhamdi}
  et~al.}{2003}]{Elmhamdi2003}
{Elmhamdi} A.,  et~al., 2003, \mn@doi [\mnras]
  {10.1046/j.1365-8711.2003.06150.x}, \href
  {http://adsabs.harvard.edu/abs/2003MNRAS.338..939E} {338, 939}

\bibitem[\protect\citeauthoryear{{Filippenko}}{{Filippenko}}{1991}]{Filippenko1991}
{Filippenko} A.~V.,  1991, in {Woosley} S.~E.,  ed., Supernovae. p.~467

\bibitem[\protect\citeauthoryear{{Fox}, {Chevalier}, {Dwek}, {Skrutskie},
  {Sugerman}  \& {Leisenring}}{{Fox} et~al.}{2010}]{Fox2010}
{Fox} O.~D.,  {Chevalier} R.~A.,  {Dwek} E.,  {Skrutskie} M.~F.,  {Sugerman}
  B.~E.~K.,   {Leisenring} J.~M.,  2010, \mn@doi [\apj]
  {10.1088/0004-637X/725/2/1768}, \href
  {http://adsabs.harvard.edu/abs/2010ApJ...725.1768F} {725, 1768}

\bibitem[\protect\citeauthoryear{{Fransson} et~al.,}{{Fransson}
  et~al.}{2014}]{Fransson2014}
{Fransson} C.,  et~al., 2014, \mn@doi [\apj] {10.1088/0004-637X/797/2/118},
  \href {http://adsabs.harvard.edu/abs/2014ApJ...797..118F} {797, 118}

\bibitem[\protect\citeauthoryear{{Gall} et~al.,}{{Gall}
  et~al.}{2014}]{Gall2014}
{Gall} C.,  et~al., 2014, \mn@doi [\nat] {10.1038/nature13558}, \href
  {http://adsabs.harvard.edu/abs/2014Natur.511..326G} {511, 326}

\bibitem[\protect\citeauthoryear{{Giuliani}}{{Giuliani}}{1982}]{Giuliani1982}
{Giuliani} Jr. J.~L.,  1982, \mn@doi [\apj] {10.1086/159939}, \href
  {http://adsabs.harvard.edu/abs/1982ApJ...256..624G} {256, 624}

\bibitem[\protect\citeauthoryear{{Hoffman}, {Leonard}, {Chornock},
  {Filippenko}, {Barth}  \& {Matheson}}{{Hoffman} et~al.}{2008}]{Hoffman2008}
{Hoffman} J.~L.,  {Leonard} D.~C.,  {Chornock} R.,  {Filippenko} A.~V.,
  {Barth} A.~J.,   {Matheson} T.,  2008, \mn@doi [\apj] {10.1086/592261}, \href
  {http://adsabs.harvard.edu/abs/2008ApJ...688.1186H} {688, 1186}

\bibitem[\protect\citeauthoryear{{Jencson}, {Prieto}, {Kochanek}, {Shappee},
  {Stanek}  \& {Pogge}}{{Jencson} et~al.}{2016}]{Jencson2016}
{Jencson} J.~E.,  {Prieto} J.~L.,  {Kochanek} C.~S.,  {Shappee} B.~J.,
  {Stanek} K.~Z.,   {Pogge} R.~W.,  2016, \mn@doi [\mnras]
  {10.1093/mnras/stv2795}, \href
  {http://adsabs.harvard.edu/abs/2016MNRAS.456.2622J} {456, 2622}

\bibitem[\protect\citeauthoryear{{Klein}, {McKee}  \& {Colella}}{{Klein}
  et~al.}{1994}]{Klein1994}
{Klein} R.~I.,  {McKee} C.~F.,   {Colella} P.,  1994, \mn@doi [\apj]
  {10.1086/173554}, \href {http://adsabs.harvard.edu/abs/1994ApJ...420..213K}
  {420, 213}

\bibitem[\protect\citeauthoryear{{Klein}, {Budil}, {Perry}  \& {Bach}}{{Klein}
  et~al.}{2003}]{Klein2003}
{Klein} R.~I.,  {Budil} K.~S.,  {Perry} T.~S.,   {Bach} D.~R.,  2003, \mn@doi
  [\apj] {10.1086/345340}, \href
  {http://adsabs.harvard.edu/abs/2003ApJ...583..245K} {583, 245}

\bibitem[\protect\citeauthoryear{{Lucy}, {Danziger}, {Gouiffes}  \&
  {Bouchet}}{{Lucy} et~al.}{1989}]{Lucy1989}
{Lucy} L.~B.,  {Danziger} I.~J.,  {Gouiffes} C.,   {Bouchet} P.,  1989, in
  {Tenorio-Tagle} G.,  {Moles} M.,   {Melnick} J.,  eds,  Lecture Notes in
  Physics, Berlin Springer Verlag Vol. 350, IAU Colloq. 120: Structure and
  Dynamics of the Interstellar Medium. p.~164, \mn@doi{10.1007/BFb0114861}

\bibitem[\protect\citeauthoryear{{Lucy}, {Danziger}, {Gouiffes}  \&
  {Bouchet}}{{Lucy} et~al.}{1991}]{Lucy1991}
{Lucy} L.~B.,  {Danziger} I.~J.,  {Gouiffes} C.,   {Bouchet} P.,  1991, in
  {Woosley} S.~E.,  ed., Supernovae. p.~82

\bibitem[\protect\citeauthoryear{{Maeda} et~al.,}{{Maeda}
  et~al.}{2013}]{Maeda2013}
{Maeda} K.,  et~al., 2013, \mn@doi [\apj] {10.1088/0004-637X/776/1/5}, \href
  {http://adsabs.harvard.edu/abs/2013ApJ...776....5M} {776, 5}

\bibitem[\protect\citeauthoryear{{Mattila} et~al.,}{{Mattila}
  et~al.}{2008}]{Mattila2008}
{Mattila} S.,  et~al., 2008, \mn@doi [\mnras]
  {10.1111/j.1365-2966.2008.13516.x}, \href
  {http://adsabs.harvard.edu/abs/2008MNRAS.389..141M} {389, 141}

\bibitem[\protect\citeauthoryear{{Ofek} et~al.,}{{Ofek}
  et~al.}{2014}]{Ofek2014}
{Ofek} E.~O.,  et~al., 2014, \mn@doi [\apj] {10.1088/0004-637X/781/1/42}, \href
  {http://adsabs.harvard.edu/abs/2014ApJ...781...42O} {781, 42}

\bibitem[\protect\citeauthoryear{{Osterbrock}}{{Osterbrock}}{1989}]{Osterbrock1989}
{Osterbrock} D.~E.,  1989, {Astrophysics of gaseous nebulae and active galactic
  nuclei}.
University science books, Mill Valley, California

\bibitem[\protect\citeauthoryear{{Pozzo}, {Meikle}, {Fassia}, {Geballe},
  {Lundqvist}, {Chugai}  \& {Sollerman}}{{Pozzo} et~al.}{2004}]{Pozzo2004}
{Pozzo} M.,  {Meikle} W.~P.~S.,  {Fassia} A.,  {Geballe} T.,  {Lundqvist} P.,
  {Chugai} N.~N.,   {Sollerman} J.,  2004, \mn@doi [\mnras]
  {10.1111/j.1365-2966.2004.07951.x}, \href
  {http://adsabs.harvard.edu/abs/2004MNRAS.352..457P} {352, 457}

\bibitem[\protect\citeauthoryear{{Sarangi}, {Dwek}  \& {Arendt}}{{Sarangi}
  et~al.}{2018}]{Sarangi2018}
{Sarangi} A.,  {Dwek} E.,   {Arendt} R.~G.,  2018, preprint, \href
  {http://adsabs.harvard.edu/abs/2018arXiv180406878S} {} (\mn@eprint {arXiv}
  {1804.06878})

\bibitem[\protect\citeauthoryear{{Smith}, {Foley}  \& {Filippenko}}{{Smith}
  et~al.}{2008}]{Smith2008}
{Smith} N.,  {Foley} R.~J.,   {Filippenko} A.~V.,  2008, \mn@doi [\apj]
  {10.1086/587860}, \href {http://adsabs.harvard.edu/abs/2008ApJ...680..568S}
  {680, 568}

\bibitem[\protect\citeauthoryear{{Smith} et~al.,}{{Smith}
  et~al.}{2009}]{Smith2009}
{Smith} N.,  et~al., 2009, \mn@doi [\apj] {10.1088/0004-637X/695/2/1334}, \href
  {http://adsabs.harvard.edu/abs/2009ApJ...695.1334S} {695, 1334}

\bibitem[\protect\citeauthoryear{{Smith}, {Silverman}, {Filippenko}, {Cooper},
  {Matheson}, {Bian}, {Weiner}  \& {Comerford}}{{Smith}
  et~al.}{2012}]{Smith2012}
{Smith} N.,  {Silverman} J.~M.,  {Filippenko} A.~V.,  {Cooper} M.~C.,
  {Matheson} T.,  {Bian} F.,  {Weiner} B.~J.,   {Comerford} J.~M.,  2012,
  \mn@doi [\aj] {10.1088/0004-6256/143/1/17}, \href
  {http://adsabs.harvard.edu/abs/2012AJ....143...17S} {143, 17}

\bibitem[\protect\citeauthoryear{{Stoll}, {Prieto}, {Stanek}, {Pogge},
  {Szczygie{\l}}, {Pojma{\'n}ski}, {Antognini}  \& {Yan}}{{Stoll}
  et~al.}{2011}]{Stoll2011}
{Stoll} R.,  {Prieto} J.~L.,  {Stanek} K.~Z.,  {Pogge} R.~W.,  {Szczygie{\l}}
  D.~M.,  {Pojma{\'n}ski} G.,  {Antognini} J.,   {Yan} H.,  2011, \mn@doi
  [\apj] {10.1088/0004-637X/730/1/34}, \href
  {http://adsabs.harvard.edu/abs/2011ApJ...730...34S} {730, 34}

\bibitem[\protect\citeauthoryear{{Utrobin} \& {Chugai}}{{Utrobin} \&
  {Chugai}}{2015}]{Utrobin2015}
{Utrobin} V.~P.,  {Chugai} N.~N.,  2015, \mn@doi [\aap]
  {10.1051/0004-6361/201424822}, \href
  {http://adsabs.harvard.edu/abs/2015A%26A...575A.100U} {575, A100}

\bibitem[\protect\citeauthoryear{{Yaron} \& {Gal-Yam}}{{Yaron} \&
  {Gal-Yam}}{2012}]{Yaron2012}
{Yaron} O.,  {Gal-Yam} A.,  2012, \mn@doi [\pasp] {10.1086/666656}, \href
  {http://adsabs.harvard.edu/abs/2012PASP..124..668Y} {124, 668}

\bibitem[\protect\citeauthoryear{{Zhang} et~al.,}{{Zhang}
  et~al.}{2012}]{Zhang2012}
{Zhang} T.,  et~al., 2012, \mn@doi [\aj] {10.1088/0004-6256/144/5/131}, \href
  {http://adsabs.harvard.edu/abs/2012AJ....144..131Z} {144, 131}

\makeatother
\end{thebibliography}

\label{lastpage}
\end{document}